\documentclass[reqno,11pt]{amsproc}

\usepackage[utf8]{inputenc}
\usepackage[round]{natbib} 
\usepackage{hyperref}
\hypersetup{
    colorlinks=true,
    linkcolor=blue,
    citecolor=blue,
    urlcolor=blue,
    }
 
\usepackage{graphicx,float,epstopdf,cite,multirow}
\usepackage{amsmath,amsfonts,amssymb,amsaddr,enumitem}
\usepackage{adjustbox,tabularx,array,booktabs,ragged2e,siunitx}
\usepackage[a4paper, margin=1in,headsep=.4in]{geometry}
\usepackage{diagbox}
\usepackage{pgfplots,pdfpages}
\usepackage{caption}
\usepackage{subcaption}

\usepackage{comment}

\newcommand{\acapotab}[1]{%
  \begin{tabular}{@{}l@{}}\strut#1\strut\end{tabular}%
}

\begin{document}

\author{Sonja Radosavljevic$^{1,*}$, Ezio Venturino$^{2,3}$, Francesca Acotto$^2$, \\ Quanli Wang$^4$, Jie Su$^4$, Alexandros Gasparatos${^4}$}
\thanks{$^*$ Corresponding author: Sonja Radosavljevic, sonja.radosavljevic@su.se  \\\hspace*{1.5em} $^1$ Stockholm Resilience Centre, Stockholm University, Albanovägen 28
SE-106 91 Stockholm, Sweden \\\hspace*{1.5em} $^2$ Dipartimento di Matematica``Giuseppe Peano'', Università di Torino, via Carlo Alberto 10,
10123 Torino, Italy \\\hspace*{1.5em}
$^3$ Laboratoire Chrono-environnement,
Université de Franche-Comt\'e,
16 route de Gray, Besan\c con, 25030, France 
\\\hspace*{1.5em} $^4$ Institute for Future Initiatives, University of Tokyo, 7-3-1 Hongo, Bunkyo-ku, Tokyo 113-0033, Japan}

\title[Sustainable intensification of SSA systems]{Sustainable intensification of small-scale aquaculture systems depends on the local context and characteristics of producers}
\date{}
\maketitle

\begin{abstract}
Aquaculture has been the fastest growing food production sector globally due to its potential to improve food security, stimulate economic growth, and reduce poverty. Its rapid development has been linked to sustainability challenges, many of which are still unresolved and poorly understood. Small-scale producers account for an increasing fraction of aquacultural output. At the same time, many of these producers experience poverty, food insecurity, and rely on unimproved production practices. We develop a stylized mathematical model to explore the effects of ecological, social, and economic factors on the dynamics of a small-scale pond aquaculture system. Using analytical and numerical methods, we explore the stability, asymptotic dynamics, and bifurcations of the model. Depending on the characteristics of the system, the model exhibits one of three distinct configurations: monostability with a global poverty trap in a nutrient-dominated or fish-dominated system; bistability with poverty trap and well-being attractors; multistability with poverty trap and two well-being attractors with different characteristics. The model results show that intensification can be sustainable only if it takes into account the local social-ecological context. In addition, the heterogeneity of small-scale aquaculture producers matters, as the effects of intensification can be unevenly distributed among them. Finally, more is not always better because too high nutrient input or productivity can lead to a suboptimal attractor or system collapse.  
\end{abstract}


\section{Introduction}

Aquaculture has been the fastest growing food production sector in the world, providing more than half of all fish for human consumption \citep{FAO2020}.  Aquaculture production, and especially small-scale aquaculture (SSA), is significant on the global level due to its potential to improve food security, stimulate economic growth, and reduce poverty in developing countries \citep{Belton}. At the same time, aquaculture development has been linked to sustainability challenges, including social issues such as inequality and common-pool resource dilemmas, and ecological issues such as eutrophication and disease outbreaks \citep{Nagel2024}. There is also a ongoing debate on who and how much benefits from participation in aquaculture, with research supporting the view that it is beneficial primarily to those who can afford it \citep{Belton}, but also to the poorest of the poor \citep{Pant}. 

An increasing fraction of aquacultural output comes from SSA producers \citep{FAO2020,Filipski}. Many of them live in regions characterized by high poverty rates, few off-farm income, employment opportunities, and high vulnerability to labor market disruptions \citep{Boughton, Kang2021}. They are often exposed to financial, climatic, and environmental risks and frequently face food insecurity, social and regulatory issues \citep{Mitra, Rahman}. SSA producers have limited knowledge of the ecology of aquaculture ponds that consist of many interdependent physical, chemical, and biological processes that are under anthropogenic and environmental influence \citep{Boyd}. Moreover, SSA producers generally lack access to improved farm technologies and rely on poor infrastructure and mechanization services. As a result, many farms are not improved, preventing SSA producers from reaching higher income and keeping them in poverty.  

The questions of small-scale aquaculture sustainability remain largely unexamined compared to other food production systems \citep{Partelow}. Most of the work is focused on conventional commercial monoculture systems, with very few exceptions that focus on small-scale aquaculture producers \citep{Little,Naylor}. Broader aquaculture research has historically focused heavily on technical aspects of production and quantitative and qualitative impact assessment \citep{Gephart2021,Henriksson,Naylor2021}. Mathematical and simulation models applied in aquaculture research usually study biophysical dynamics and explore biotic and abiotic factors that affect fish growth and the behavior of ecological populations. These include the dynamic energy budget model \citep{Koojiman}, the thermal growth coefficient model \citep{Jobling}, the biomass-based models \citep{Svirezhev}, or individual-based models \citep{Lu}. In other cases, aquaculture models focus on bioeconomic dynamics \citep{Nobre}, optimization problems \citep{Kvamsdal}, or the effects of climate change \citep{Varga}. 

Unlike agriculture research, where dynamical system models have been used to explore social-ecological interactions and poverty traps \citep{Lade, Radosavljevic2020, Radosavljevic2021,Sanga2024}, models in aquaculture rarely adopt a social-ecological systems approach \citep{Levin2013} and lack the ability to explore rural poverty and development pathways from a holistic point of view \citep{Bene2016}. An exception is the work of \citet{Filipski} in which the authors use the general equilibrium model to study the effects of small-scale commercial aquaculture on poverty. Due to this research gap, little is known about how intertwined social and ecological processes shape the dynamics of a SSA system and, in turn, affect poverty and food security of SSA producers.

The purpose of this paper is two-fold. We aim to explore the long-term dynamics of small-scale pond aquaculture systems created by intertwined social-ecological processes. We also aim to identify the leverage points within the system where interventions could be useful and to identify critical points where shocks could be dangerous. To this end, we adopt a social-ecological systems approach and use empirical and theoretical knowledge to develop a stylized model that represents a small-scale pond aquaculture system. 

The most useful framework for achieving our goal is the use of continuous-time dynamical system theory. By developing a dynamical system model based on systems of ordinary differential equations, we can employ analytical and numerical methods to explore the short- and long-term behavior of the model.  Understanding the real system and the possibilities for interventions hinges on knowing the asymptotic behavior of the model, that is, the number, location, and properties of equilibrium points \citep{Radosavljevic2023}. Stable equilibrium points, the so-called attractors, represent the states of the system that persist forever if the system is undisturbed. A basin of attraction is a set of states that converge to the same attractor. 

Conceptually, we use the notion of the poverty trap and well-being to distinguish between unwanted and wanted states of the SSA system. A poverty trap is an unwanted state of social-ecological systems formed by self-reinforcing mechanisms that keep individuals in persistent poverty \citep{Barrett2013, Haider2018}. In dynamical system models, poverty traps are identified with unwanted attractors, while well-being states correspond to wanted attractors. Management and development efforts typically aim to ``push" the system into the well-being basin of attraction and toward the well-being attractor \citep{Barrett2016}. 

The model developed here represents a small-scale pond aquaculture system. It combines the ecological dynamics of fish biomass growth and the nutrient life cycle, the economic dynamics of assets, and the social-ecological interactions that connect the system variables. The model allows us to explore under what conditions the intertwined dynamics of assets, fish biomass, and nutrients lead to different system-wide outcomes, such as well-being or permanent poverty. We rely on understanding asymptotic and transient behavior to identify leverage points and suitable interventions.

The paper is organized as follows. In Section 2 we present the model of a small-scale pond aquaculture system, including its empirical and theoretical assumptions. Section 3 contains the main analytical results related to the model equilibrium points and the conditions for their feasibility and stability. Further mathematical details can be found in the Appendix.
The numerical results of the stability and bifurcation analysis are in Section 4. Finally, we discuss the results and their social-ecological significance, implications for management, open questions, and future research.


\section{A stylized small-scale pond aquaculture model}

This section has two objectives. First, we describe
the conceptual model that underlies the mathematical
model. Second, we develop the mathematical
model using a system of nonlinear ordinary differential equations. 

\subsection{The conceptual model}

We base our causal understanding of pond aquaculture systems on empirical studies such as \citep{Belton, DeSilva, Filipski, F4L}, and the first-hand experience of three of the authors conducting small-scale aquaculture research in developing countries such as Myanmar, Egypt, and Bangladesh \citep{DamLam, Dompreh, Rossignoli2023a, Rossignoli2023b, Wang2023, Wang2024}. In short, small-scale aquaculture is an activity that produces fish for household consumption and sales in the market. In this sense, it can contribute to household nutrition and income generation, having a positive outcome for food security and livelihood. Small-scale producers are generally poor and face food insecurity. Adopting fish culture or increasing the technical efficiency of existing fish production can increase levels of income and fish consumption, and consequently reduce producers' poverty and food insecurity. Maintaining the required water quality and food supply for the farmed species is necessary to obtain high yields and high quality products. However, the ecology of aquaculture ponds involves many interdependent physical, chemical and biological processes \citep{Boyd}. In addition to being complex, the biological and biophysical characteristics of small-scale ponds are open to environmental and anthropogenic influences that operate at different temporal, spatial, and organizational levels. Tools from ecology can be very useful for modeling these dynamics. Representing all of them in a single model would be too complicated and also unnecessary to answer our research questions. Thus, we abstract from a detailed description of socioeconomic, biochemical, or ecological processes in the system and attempt to capture its qualitative behavior. The conceptual model is based on the following assumptions:
\begin{enumerate}
    \item increased aquaculture production increases SSA producers' income;
    \item SSA producers' income has positive effects on fish growth (e.g., by providing fish feed);
    \item fish growth has positive effects on fish production;
    \item nutrients have a positive effect on fish growth, but high levels of nutrients harm water quality, which, in turn, can decrease fish growth and increase mortality.
\end{enumerate}

\begin{figure}[h]
\centering
\includegraphics[width=0.8\linewidth]{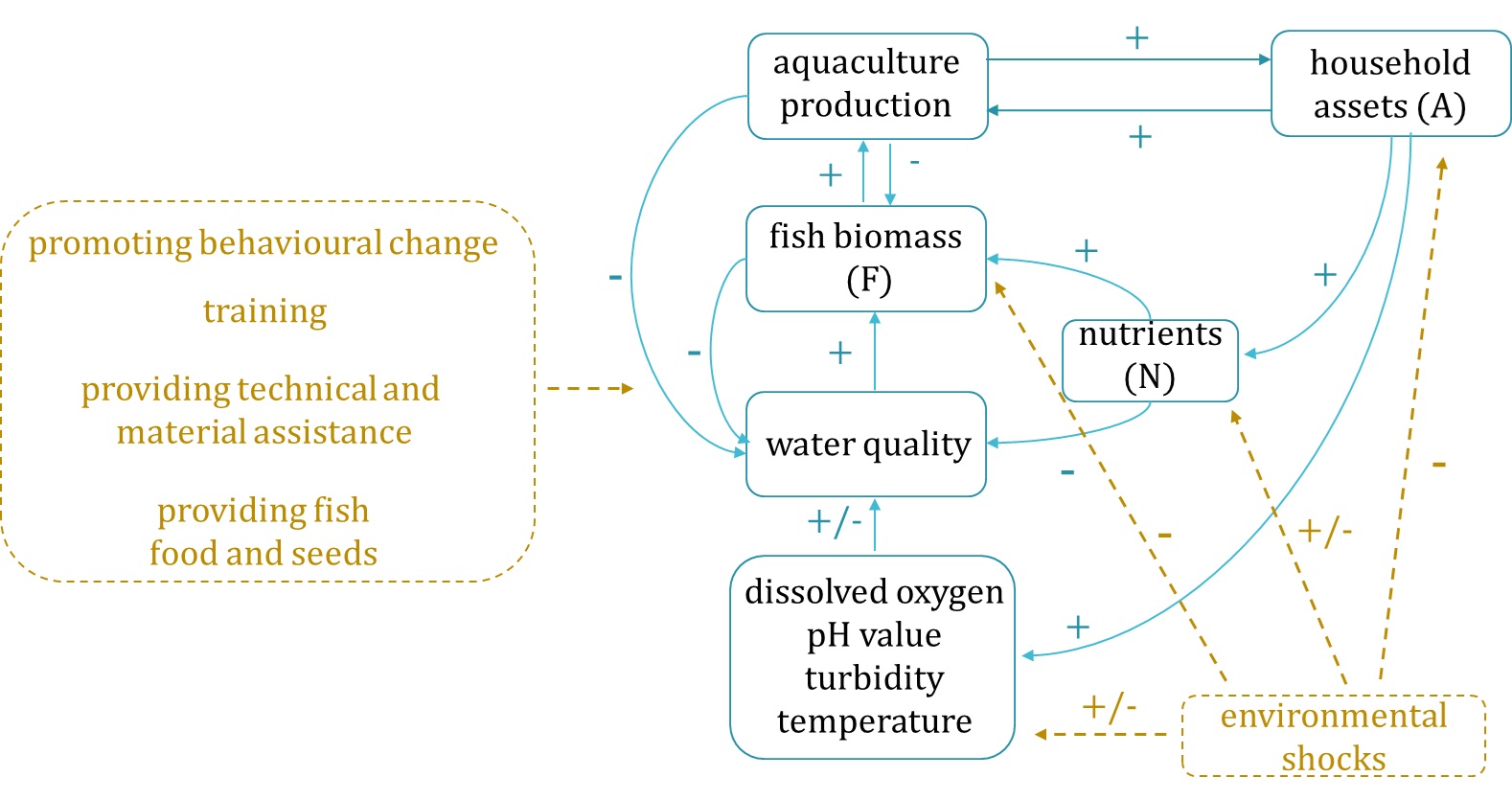}
\caption{\footnotesize Causal loop diagram for the stylized pond aquaculture model. Blue arrows represent long-term processes in the system. External interventions and shocks are given in olive-green. Dashed olive-green arrows represent short-term processes.}
\label{fig:cld}
\end{figure}

The conceptual model is visualized using a causal loop diagram in Figure \ref{fig:cld}. It represents the long-term dynamics of the system created through intertwined biochemical, economic, and social-ecological processes, such as fish growth, nutrient input, uptake by fish, and loss, and aquaculture production. External interventions and shocks (e.g., environmental or financial) are seen as short-term processes that do not change the way the system works but change the initial conditions. 

\subsection{The mathematical model}

Based on the conceptual model and Figure \ref{fig:cld}, we select the state variables for the mathematical model. These are: household assets, $A$, fish biomass, $F$, and nutrients, $N$. The choice of assets and fish biomass is not surprising, as they are needed for aquaculture, but choosing nutrients instead of water quality might need an explanation. Water quality is hard to define from a mathematical perspective if we want to keep simplicity and include realism in the model. It has a multidimensional nature that includes water temperature, pH, dissolved oxygen, and turbidity. Many of these dynamics are complex and driven by influences beyond the producers' control. On the other hand, nutrient dynamics is heavily impacted by human activities, such as agriculture or choice of fish feed. Increasing nutrients in fishponds usually enhances fish growth and survival. However, increasing the amount of nutrients above a certain level reduces the quality of the water, which can have negative effects on the growth and survival of fish. Modeling nutrients explicitly, therefore, allows us to explore how farmers' decisions and practices affect the dynamics of the system and include, in a roundabout way, water quality.


\emph{Assets dynamics.} We extend the classical Solow model \citep{Barro}, where productivity depends only on assets, and assume that it depends on assets and fish biomass. The production function is, therefore, a Cobb-Douglas production function of the form
$f(A,F)=bA^{\alpha}F^{\beta}$.
The parameter $b>0$ denotes the productivity factor and reflects the knowledge, practices, and technology of SSA producers. According to \citep{Asamoah}, small- and medium-scale aquaculture producers exhibit a constant or increased return to scales with elasticity coefficients $\alpha + \beta \ge 1$. 
Following \citep{Kraay}, we assume that the savings rate is an S-shaped function defined by $s(A)=s\frac{A^2}{p+A^2}$. 
Taking these considerations into account, the rate of change of household assets is given by the following equation:
\begin{align}\label{assets}
    \frac{dA}{dt} &= bs \frac {A^2} {p+A^2} A^{\alpha}F^{\beta} - q A,
\end{align}
where the negative term $-qA$ denotes asset depreciation rate.

\emph{Fish dynamics.} The classical bioeconomic model \citep{Clark} assumes that fish population growth follows the logistic equation 
\begin{equation*}
    \displaystyle\frac{dF}{dt}=rF-mF-c F^2-hF,
\end{equation*}
where the positive term $rF$ denotes fish biomass growth, and the negative terms $-mF$ and $-hF$ denote mortality and harvest proportional to the fish biomass, respectively. The negative term $-cF^2$ represents intraspecific competition for resources. 

Nutrients play an important role in shaping fish growth dynamics, but the relationship between nutrient concentration in water and fish growth is complex. Nutrients are necessary for fish growth, and increasing their concentration to a certain level increases fish growth. However, too high nutrient levels can harm water quality and, in turn, decrease fish growth rate and increase fish mortality. To model these opposing effects of nutrients on fish growth, the effect of nutrients on fish growth rate is given by $r(N)=ru\frac{N}{v+N^2}$. Thus, the classical logistic fish growth equation is modified as follows:
\begin{align}\label{fish}
     \frac{dF}{dt} &= r u\frac{NF}{v+N^2} - mF- c F^2 - h F .
\end{align}

\emph{Nutrient dynamics.} Nutrients enter the pond water from two main sources: agricultural run-off at a constant input, $k$, or fish feed, expressed as $g\frac{AF}{z+A}$. The functional form implies that the nutrient input is limited by the fish biomass and the farmers' assets. For low asset levels, the input of nutrients is low and limited by the available assets, but for a high asset level, the nutrient amount is proportional to the fish biomass and limited by the amount of fish feed needed. The uptake of nutrients by fish is proportional to the fish biomass and is expressed as $-u \frac{NF} {v+N^2}$. Nutrient loss due to natural processes is given by $-\ell N$. These considerations lead us to the following equation for the nutrient dynamics:
\begin{align}\label{nutrients}
    \frac{dN}{dt} &= k + g \frac {AF} {z+A} - u \frac {NF} {v+N^2} - \ell N.
\end{align}

Combining equations (\ref{assets})-(\ref{nutrients}), we come to the model of SSA system:
\begin{equation}\label{model}
    \begin{aligned}
    \frac{dA}{dt} &= bs \frac {A^2} {p+A^2} A^{\alpha}F^{\beta}- q A, \\
   \frac{dF}{dt} &= r u\frac{NF}{v+N^2} - m F - c F^2 - h F, \\
    \frac{dN}{dt} &= k + g \frac {AF} {z+A} - u \frac {NF} {v+N^2} - \ell N.
    \end{aligned}
\end{equation}
All model parameters, their meanings, values, or range of values, are specified in Table ~\ref{Table:param}.

\begin{table}[h]\footnotesize
\begin{tabularx}{\textwidth}{p{2cm} p{7cm} p{2cm}  p{3.5cm}}
\toprule
\bf{Parameters} & {\bf Interpretation} & {\bf Values} & {\bf References}
\\\midrule
$b$ & Factor of productivity & $[0.1,1.5]$ & \citep{Asamoah}\\
$s$ & Maximal assets savings rate & $1$ & Assumed\\
$p$ & Half-saturation point of assets savings rate & $10$ & Assumed\\
$q$ & Assets depreciation rate  & $[0.05,0.5]$ & \citep{Nobre}\\
$\alpha$, $\beta$ & Elasticity coefficients &  $\alpha+\beta\ge 1$ & \citep{Asamoah} \\
$r$ & Fish growth factor & $[0.002,0.008]$ & \citep{Scheffer}\\
$u$ & Nutrient uptake rate & $[0.1,1]$ & Assumed  \\
$v$ & Square of the optimal nutrient concentration   & $10$ & Assumed\\
$m$ & Fish mortality rate & $[0.002,0.01]$ & \citep{Scheffer} \\
$c$ & Fish competition rate & $[0.001,0.002]$  & Assumed \\
$h$ & Fish harvest rate & $h\geq 0$ & Assumed  \\
$k$ & Constant run-off nutrient input & $[0,1]$ & Assumed \\
$g$ & Fish feed nutrient input rate & $ [0,1]$ & Assumed \\
$z$ & Half-saturation point of nutrient input rate & $5$ & Assumed \\
$\ell$ & Natural nutrient loss rate & $0.1$ & Assumed  
\\ \bottomrule
\end{tabularx}
\smallskip
\caption{List of parameters used in model \eqref{model}.}
\label{Table:param}
\end{table}


\section{Results: analytical investigation of the model} \label{analysis}

By developing a dynamical system model, we make use of two mathematical techniques: analysis of stability and bifurcations. Stability analysis studies the asymptotic properties of solutions and explores what happens with the system after a long time. Bifurcation analysis allows us to investigate how changes in parameter values lead to qualitatively different behavior of the model. In this section, we focus on the stability study from an analytical point of view, after discussing the feasibility of equilibrium points. Bifurcations are investigated numerically in the next section. 

The model (\ref{model}) is nonlinear and fairly difficult to explore analytically. To enable some of the analytical methods for stability investigation, we simplify the model assuming the maximal assets savings rate and the elasticity coefficients equal to one, i.e.,  $s=\alpha=\beta=1$. With this assumption, the model reads as follows:
\begin{equation}\label{modelSimple}
    \begin{aligned}
    \frac{dA}{dt} &= b \frac {A^2} {p+A^2} AF- q A, \\
   \frac{dF}{dt} &= r u\frac{NF}{v+N^2} - m F - c F^2 - h F, \\
    \frac{dN}{dt} &= k + g \frac {AF} {z+A} - u \frac {NF} {v+N^2} - \ell N.
    \end{aligned}
\end{equation}

The following two subsections are devoted to the analytical study of this simplified model, in particular, its equilibrium points' feasibility and local stability. The main results are summarized in Table \ref{tab_eq}. Some mathematical details are given in the Appendix.

\subsection{Feasibility of equilibrium points}

The simplified model (\ref{modelSimple}) allows only the following three equilibria:
$$
E_1= \left( 0,0, \frac k{\ell}\right), \quad E_2 = \left( 0, F_2, N_2\right), \quad E_*=(A_*, F_*, N_*).
$$
The equilibrium with only nutrients, $E_1$, is explicitly known and unconditionally feasible. Instead, the nonlinear system (\ref{modelSimple}) is too complex to explicitly determine the components of the equilibrium without assets, $E_2$, and those of the coexistence equilibrium, $E_*$. However, we can look for sufficient conditions for their feasibility.

In the assets-free case, we can look for conditions that ensure at least one graphical intersection point, in the first quadrant of the $N$-$F$ plane, between two curves we find from the equilibrium equations. In the coexistence case, similarly, we are interested in intersecting three surfaces in the first octant of the $A$-$N$-$F$ space. The sufficient conditions for the feasibility of $E_2$ and $E_*$ given in Table \ref{tab_eq} are obtained in the Appendix.

\subsection{Local stability of equilibrium points}

The Jacobian of the model (\ref{modelSimple}) is
\begin{equation}\label{Jac}
J = \left[
\begin{array}{cccc}
J_{11} & \frac {bA^3}{p+A^2} & 0 \\
0 & \frac {ruN}{v+N^2} -(m+h) - 2cF & ruF \frac {v-N^2}{(v+N^2)^2} \\
gF \frac z{(z+A)^2} & \frac {gA}{z+A} - \frac {uN}{v+N^2} & J_{33}
\end{array}
\right],
\end{equation}
where 
\begin{equation*}
    \begin{aligned}
J_{11} &= bF \frac {3A^2 (p+A^2) - 2 A^4}{(p+A^2)^2} -q , \quad
J_{33} = - uF \frac {v-N^2}{(v+N^2)^2} - \ell.
\end{aligned}
\end{equation*}

\subsubsection{Nutrients-only point, $E_1$}

By evaluating (\ref{Jac}) at equilibrium $E_1$, a lower triangular matrix is obtained, from which
the eigenvalues are easily found. They are 
$$
-q, \quad -f, \quad \frac {ruk \ell}{v \ell^2 + k^2 } - (m+h).
$$

Thus, we have the following local asymptotic stability condition:
\begin{equation}\label{E1_stab}
ru k \ell < (m+h) (v \ell^2 + k^2).
\end{equation}

\subsubsection{Assets-free point, $E_2$}

At $E_2$ the characteristic equation factorizes and one eigenvalue is explicitly found, that is, $-q$. This eigenvalue is always negative, thus it does not affect the local asymptotic stability of the assets-free equilibrium point.

We use the notation $J_{[m,n]}$ for the submatrix of $J$ in which the rows and columns $m$ and $n$ are
preserved.
For the remaining $2 \times 2$ minor, $J_{[2,3]}(E_2)$, we use the Routh-Hurwitz criterion.
The determinant of this minor is 
$$
\det (J_{[2,3]}(E_2)) = c F_2 \left[  \ell + u F_2 \frac {v-N_2^2}{(v+N_2^2)^2} \right]
+ ruF_2 \frac {v-N_2^2}{(v+N_2^2)^2} \frac {uN_2}{v+N_2^2}.
$$
Its trace, instead, using the second equilibrium equation, reduces to
$$
{\textrm {tr}}(J_{[2,3]}(E_2)) = - c F_2 - \ell - u F_2 \frac {v-N_2^2}{(v+N_2^2)^2} <0.
$$

Thus, the Routh-Hurwitz condition on the determinant gives the following local asymptotic stability condition:
\begin{equation}\label{E2_stab}
c F_2 \left[ \ell + \frac {uv F_2 }{(v+N_2^2)^2} \right]
+ \frac {ruv F_2 }{(v+N_2^2)^2} \frac {uN_2}{v+N_2^2}
> \frac {ruF_2 N_2^2}{(v+N_2^2)^2} \frac {uN_2}{v+N_2^2}
+ \frac {cu F_2^2 N_2^2}{(v+N_2^2)^2} .
\end{equation}

\subsubsection{Coexistence, $E_*$}

In the coexistence case, we can find the local asymptotic stability condition using the Routh-Hurwitz criterion for a cubic equation, i.e., $RH3(J(E_*))$. However, an explicit determination of $RH3(J(E_*))$ is too much involved and will not shed more light on the problem, so we do not analytically investigate them further. We explore the coexistence point stability using numerical methods in the following section.

\begin{table}[h]
\footnotesize
\begin{tabularx}{\textwidth}{p{3cm} p{6cm}  p{6cm}}\\
\toprule
\bf{Equilibria} & \bf{Feasibility} & \bf{Local Stability} \\ 
\toprule
$E_1= \left( 0,0, \frac k{\ell}\right)$ & unconditionally feasible & asymptotically stable iff \eqref{E1_stab} \\
\midrule
$E_2 = \left( 0, F_2, N_2\right)$ & 
\acapotab{not feasible if \eqref{Psi_feas}, with \eqref{Theta_deriv2_positive}\\ or \eqref{N**}, when \eqref{E2_no}\\ \\ feasible and unique if \eqref{Psi_feas},\\ with \eqref{Theta_deriv2_positive} or \eqref{N**}, when \eqref{E2!} \\ \\ saddle-node if \eqref{Psi_feas}, with \eqref{Theta_deriv2_positive}\\ or \eqref{N**}, when \eqref{E2-saddle} and \eqref{saddle-condition}}
 & asymptotically stable iff \eqref{E2_stab} \\
\midrule
$E_*=(A_*, F_*, N_*)$ &  see Tables \ref{tab_a1}, \ref{tab_a2}, \ref{tab_a3}, \ref{tab_b1}, \ref{tab_b2}, and \ref{tab_b3} & asymptotically stable iff $RH3(J(E_*))$ \\
\bottomrule
\end{tabularx}
\smallskip
\caption{Equilibria of model \eqref{modelSimple}: feasibility and local stability conditions.}
\label{tab_eq}
\end{table}


\section{Numerical results and implications}

The purpose of this section is to use stability analysis to explore the long-term dynamics of the model. In structurally unstable systems, the dynamics can change qualitatively when the parameters pass threshold values. Knowing the properties of system dynamics if the parameters change or in the case of uncertainty in the values of the parameters improves understanding of the system and management decisions. Therefore, we use bifurcation analysis to explore qualitative changes in the system structure for a range of parameter values.

\subsection{Bistability and multistability} 

The analysis in Section 3 shows that, depending on parameter values, the model (\ref{modelSimple}) can have one, two, or more stable equilibrium points that represent the nutrients-only, the assets-free, and the coexistence states.
For the complete model (\ref{model}), we use numerical analysis to discover the attractors and characterize the basins of attraction, as shown in Figures \ref{fig:twoEP} and \ref{fig:threeEP}.

In social-ecological terms, the assets-free state, $E_2$, is an unwanted equilibrium because it represents the collapse of aquaculture production due to the lack of assets. In poverty trap research, this attractor is a poverty trap. In contrast to this, the coexistence equilibria, denoted here as $E_*$ and $E_{**}$, represent the well-being attractors of the aquaculture system, i.e., the situations when the system functions more or less successfully. The efficacy of the system is assessed by comparing the location of coexistence equilibria in the phase space. Similarly, the resilience to shocks is assessed by estimating the distance between the equilibrium point and the boundary of its basin of attraction. 

\subsubsection{Bistability}
\begin{figure}[h]
\centering
\includegraphics[width=0.8\linewidth]{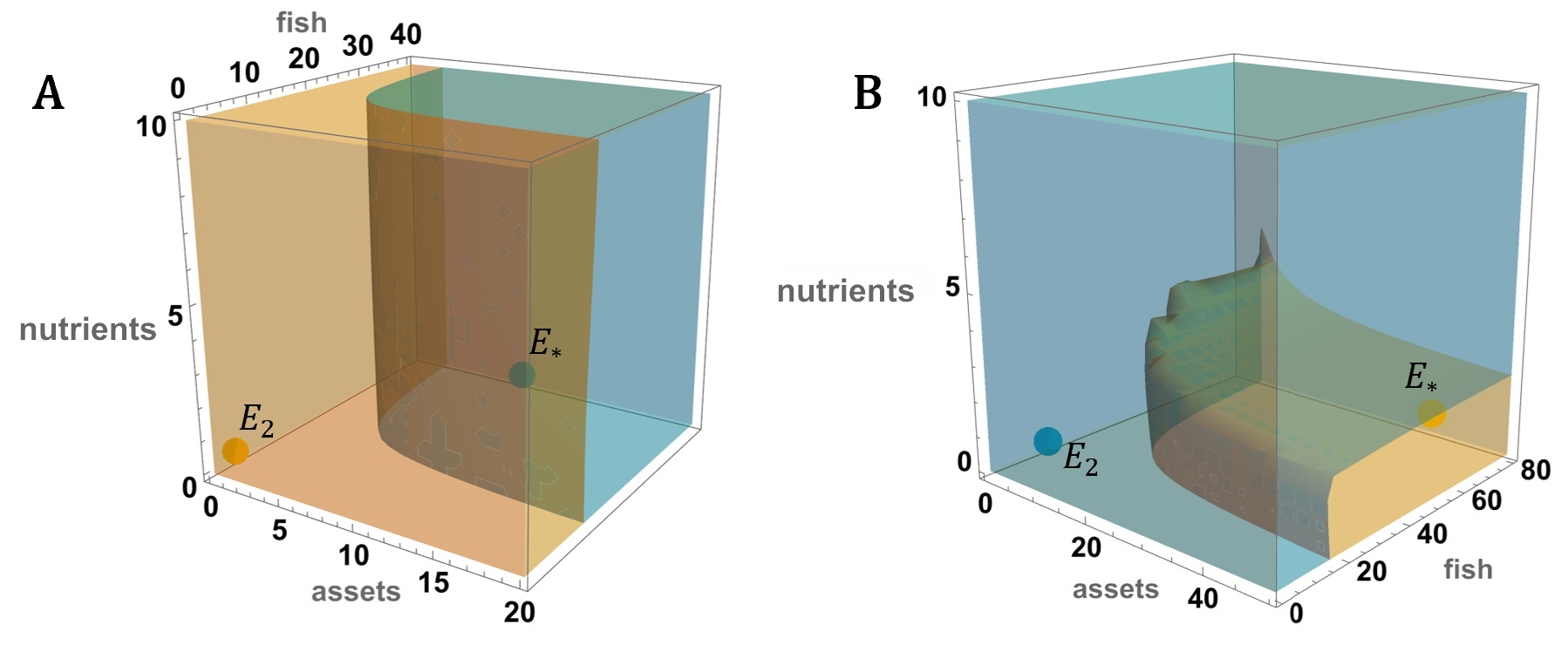}
\caption{\footnotesize The model (\ref{model}) is bistable for various combinations of the parameter values. (A) In the left panel, we consider $b=0.3,$ $p=10,$ $\alpha=0.5,$ $\beta=0.5,$ $q=0.4,$ $r=0.9,$ $u=0.3,$ $m=0.01,$ $c=0.002,$ $k=0.1,$ $g=0.1,$ and $\ell=0.1$. 
(B) In the right one, we set $b=0.3,$ $p=50,$ $\alpha=0.5,$ $\beta=0.5,$ $q=0.4,$ $r=0.9,$ $u=0.3,$ $m=0.01,$ $c=0.001,$ $k=0.5,$ $g=0.1,$ and $\ell=0.1$. Assets-free equilibrium, $E_2$, represents the undesired poor state and coexistence equilibrium, $E_*$, is the well-being state.}
\label{fig:twoEP}
\end{figure}

Figures \ref{fig:twoEP}A and \ref{fig:twoEP}B show bistable aquaculture systems, but their behavior is quite different. The shape of the poor basin of attraction in Figure \ref{fig:twoEP}A indicates that nutrient levels do not play a role in its characterization. Regardless of the level of nutrients, the trajectories will converge toward collapse equilibrium, $E_2$, if the levels of assets and fish biomass are low. However, if the assets and the biomass of the fish are sufficiently high, the trajectories will converge to the well-being attractor, $E_*$. In contrast, Figure \ref{fig:twoEP}B implies the importance of nutrients in characterizing the basins of attraction and the complex relationship between the state variables. Judging by the relative size of the basins of attraction, we see that the well-being state, $E_*$, is less resilient than the collapse state, $E_2$. The well-being state can be reached only if the initial conditions meet the requirements of having sufficient assets and fish biomass, and low nutrients. If any of the conditions fails, the trajectory converges toward the collapse state $E_2$.

In the scenario in Figure \ref{fig:twoEP}A, the collapse equilibrium, $E_*$, and the well-being equilibrium, $E_{**}$, are resilient to changes in nutrient values. Depending on the location of its initial conditions within the basin of attraction, the trajectory could be pushed toward the well-being state by increasing the levels of assets, fish biomass, or both. Similarly, the well-being equilibrium, $E_{**}$, could be disturbed by shocks that decrease the levels of assets, fish biomass, or both. 

The scenario in Figure \ref{fig:twoEP}B shows another complexity: a higher input rate for runoff nutrients and a higher half-saturation of the savings rate created a situation in which nutrients play an important role. The resilience of the well-being attractor is reduced in comparison to the scenario in Figure \ref{fig:twoEP}A and can be affected by increasing nutrients. At the same time, the resilience of the poverty trap is increased, and disturbing it may require a simultaneous reduction in nutrients and an increase in assets and fish biomass.

\subsubsection{Multistability} 

\begin{figure}[h]
\centering
\includegraphics[width=0.8\linewidth]{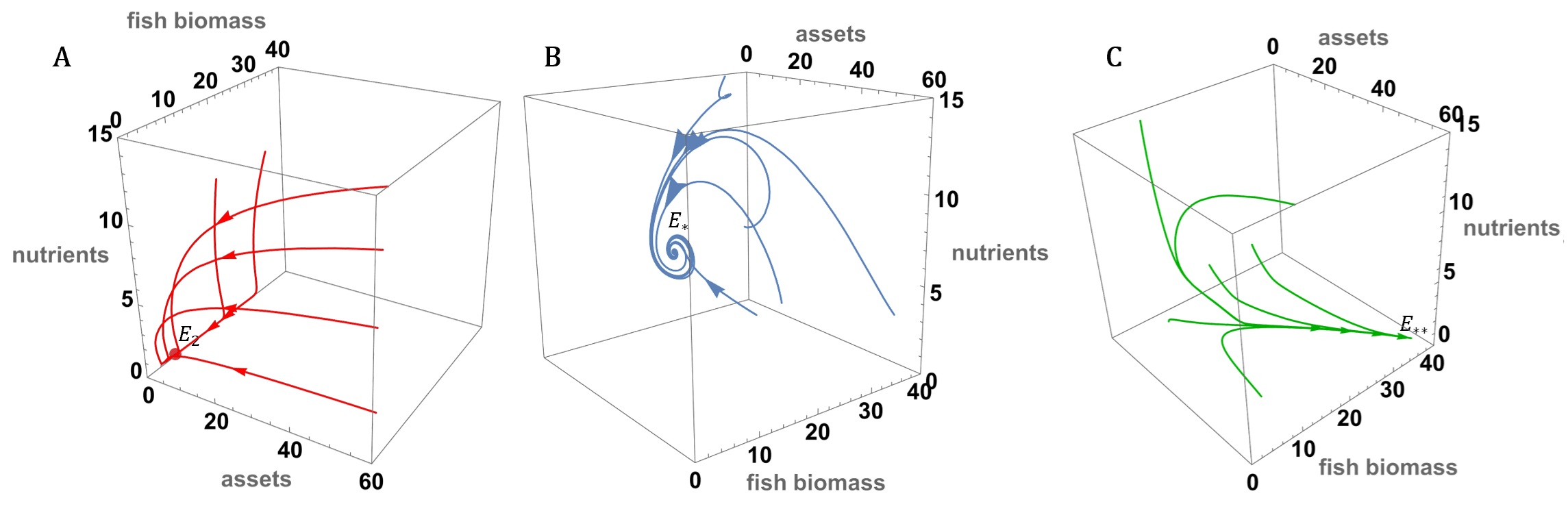}
\caption{\footnotesize Model (\ref{model}) has three stable states for $b=0.5,$ $p=10,$ $\alpha=0.5,$ $\beta=0.5,$ $q=0.4,$ $r=0.9,$ $u=0.3,$ $m=0.01,$ $c=0.001,$ $k=0.1,$ $g=0.1,$ and $\ell=0.1$. Trajectories that converge toward the poor state (assets-free equilibrium, $E_2$) are in red, while blue and green trajectories are those that converge toward the well-being states (coexistence equilibria, $E_*$ and $E_{**}$).}
\label{fig:threeEP}
\end{figure}

Increasing the productivity from $b=0.3$ in Figure \ref{fig:twoEP} to $b=0.5$ resulted in the appearance of three stable equilibria, i.e., $E_2$, $E_*$, and $E_{**}$, in Figure \ref{fig:threeEP}.

Figure \ref{fig:threeEP}A suggests that the basin of attraction of the assets-free equilibrium, $E_2$, consists of initial states that can have all nutrient values provided a low level of assets or low level of fish biomass. The basin of attraction of $E_*$ comprises initial states with sufficiently high values of all state variables, see Figure \ref{fig:threeEP}B. The most unusual and potentially counterintuitive basin of attraction is that of $E_{**}$. It is characterized by initial conditions with low nutrient value and all, except very low levels of assets and fish biomass. It can also be reached for all nutrient values provided an intermediate level of fish biomass and almost all levels of assets, see Figure \ref{fig:threeEP}C. 

Comparing the positions of $E_*$ and $E_{**}$ in the phase space (Figures \ref{fig:threeEP}B and \ref{fig:threeEP}C) leads to the conclusion that $E_{**}$ is preferred to $E_*$ because it has higher levels of assets and fish biomass. The trajectories converging towards $E_{**}$ can be disturbed by increasing nutrient levels, which would force them toward $E_*$. Another possibility is drastically reducing the level of assets or fish biomass and forcing trajectories toward $E_2$. Risks in the system can come in different forms, but do not have to lead to catastrophic outcomes. 
Using the same reasoning, we can assess the efficiency of management strategies and interventions. To be efficient, a poverty alleviation strategy should increase the assets or fish biomass and probably reduce the nutrients in the system.

\subsection{Bifurcations}

The derivation in Section 3 showed that the analytical results are hard to obtain even for the simplified model (\ref{modelSimple}). The parameters $\alpha$ and $\beta$ were eliminated in the simplified model (\ref{modelSimple}) by setting them to $1$. However, they are very important for understanding the consequences of social-ecological interactions related to the use of new technology. Changes in the values of these parameters can be understood as the adoption of innovation and training.

Similarly, the parameters $k$ and $g$ define the amount of nutrients that enter the system, unintentionally as runoff (in the case of $k$) or intentionally through fish feed (in the case of $g$). These parameters can be related to SSA producers' decisions and the capacity to manage agricultural and aquacultural nutrients and to choose the type and amount of fish feed. Parameters $r$, $u$, and $v$ describe ecological processes of growth and nutrient uptake and depend on species, local ecological conditions, and the type of fish feed. Finally, $b$, $p$, and $q$ help define asset dynamics and can be related to SSA productivity, assets savings rate and depreciation rate, respectively. 

In the remainder of the section, we present the results of the bifurcation analysis for some of these parameters.

\subsubsection{Productivity parameter $b$}

\begin{figure}[h]
\centering
\includegraphics[width=0.4\linewidth]{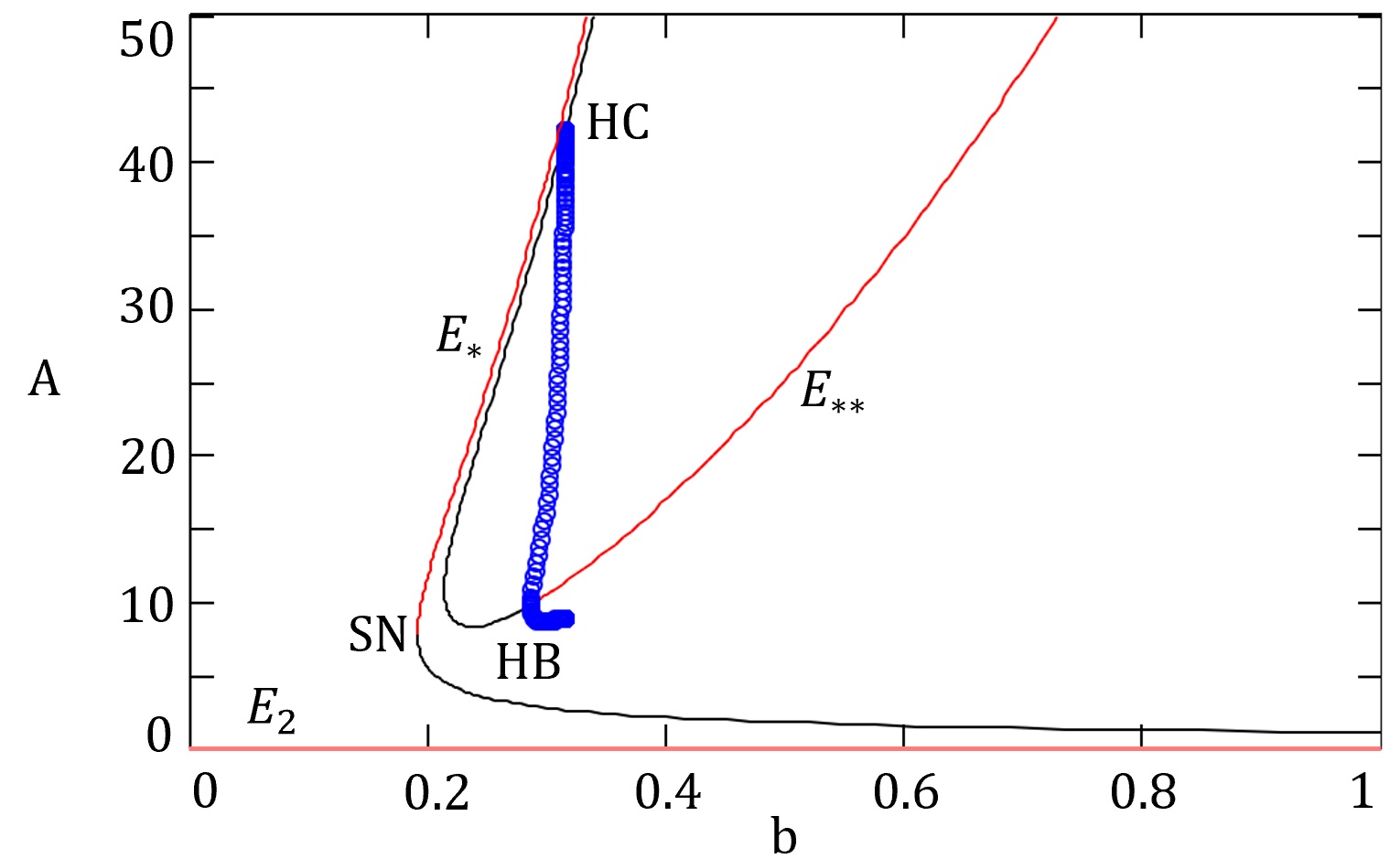}
\caption{\footnotesize Bifurcation analysis shows the existence of one, two, or three stable equilibrium points for different values of the parameter $b$. Red lines denote stable equilibrium points. Black lines denote unstable equilibria. Blue circles are unstable limit cycles. $E_2$ is the branch of assets-free equilibria. $E_*$ and $E_{**}$ are branches of coexistence equilibria. Saddle-node, Hopf, and homoclinic bifurcations are denoted by $SN$, $HB$, and $HC$, respectively. The parameter values are $p=10,$ $\alpha=0.5,$ $\beta=0.5,$ $q=0.4,$ $v=2,$ $z=5,$ $r=0.9,$ $u=0.3,$ $m=0.01,$ $c=0.001,$ $k=0.1,$ $g=0.1,$ and $\ell=0.1$.}
\label{fig:Model3D_b}
\end{figure}

The bifurcation diagram in Figure \ref{fig:Model3D_b} shows changes in the number and stability of the equilibrium points when the parameter $b$ is varied.
The assets-free state, $E_2$, exists and is stable for all parameter values, reminding us that system collapse is always an option. Increasing the productivity parameter, $b$, creates the second stable equilibrium point, $E_*$, through a saddle-node bifurcation. Increasing $b$ even more leads to a Hopf bifurcation and the appearance of the third stable equilibrium point, $E_{**}$. The resilience of $E_*$ is reduced and the well-being attractor becomes more sensitive to changes in asset values. In contrast to this, the shape of the basin of attraction of $E_{**}$ indicates the resilience of this attractor to changes in asset values.

\subsubsection{Fish feed nutrients}

\begin{figure}[h]
\centering
\includegraphics[width=0.4\linewidth]{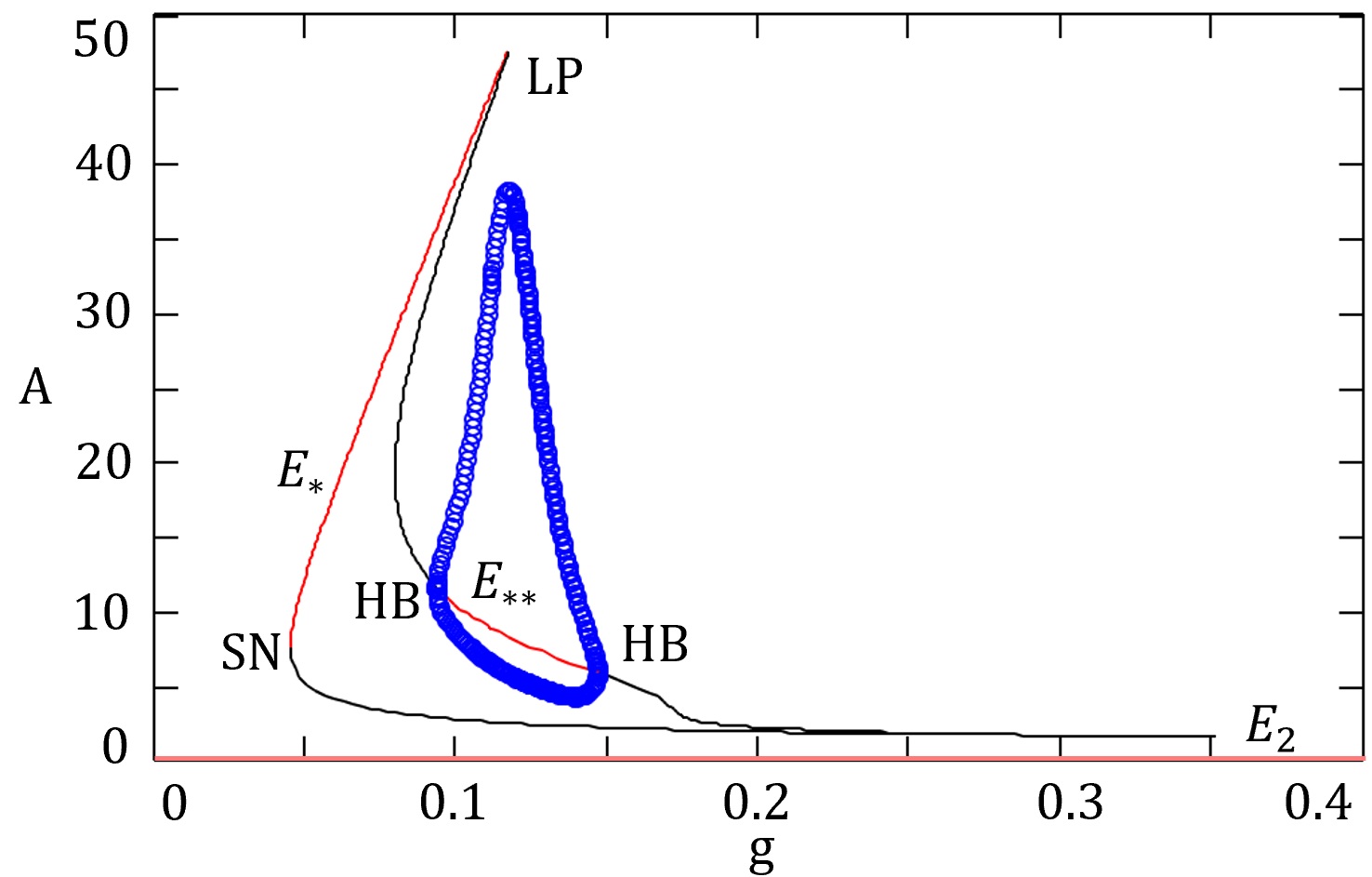}
\caption{\footnotesize Bifurcation analysis shows the existence of one, two, or three stable equilibrium points for different values of the parameter $g$. Saddle-node and Hopf bifurcations are denoted by $SN$ and $HB$, respectively. $LP$ stands for a limit point. The parameter values are $b=0.3,$ $p=10,$ $\alpha=0.5,$ $\beta=0.5,$ $q=0.4,$ $v=2,$ $z=5,$ $r=0.9,$ $u=0.3,$ $m=0.01,$ $c=0.001,$ $k=0.1,$ and $\ell=0.1$.}
\label{fig:Model3D_g}
\end{figure}

Taking the fish feed nutrients, $g$, as a bifurcation parameter leads to the bifurcation diagram in Figure \ref{fig:Model3D_g}. Assets-free equilibrium, $E_2$, exists for all values of $g$, which implies that the collapse of the aquaculture system is always possible. Collapse is unavoidable for either too low or too high values of $g$ with $E_2$ as the only attractor. In the former case, the collapse is caused by a low fish growth rate created by a lack of nutrients. In the latter case, the system collapses due to the detrimental effects of nutrients on water quality and, in turn, on fish growth. For intermediate values of the input rate of nutrients ($g\in \lbrack 0.045,0.094 \rbrack$), the second stable state, $E_*$, emerges after a saddle-node bifurcation. After a Hopf bifurcation at $g=0.094$, the third stable state $E_{**}$ appears. As in the case where the productivity parameter $b$ is increased, $E_{**}$ is less preferred to $E_*$.

\section{Discussion}

\subsection{Structural understanding}
The model developed in this paper is based on social-ecological systems research \citep{Levin2013} and has a strong focus on poverty traps \citep{Barrett2013,Barrett2016,Haider2018}. 
Previous models of poverty traps have been used as exploratory tools to provide insight into the dynamics of a system, test hypotheses, develop scenarios and answer the ``what if" questions \citep{Banitz22, Eppinga2024}. The models were mainly conceptual, rooted in the neoclassical economic tradition \citep{Barro}, and discussed economic causes and solutions to poverty \citep{Barrett2016,Blume2020}. In recent years, multidimensional poverty trap models of agricultural systems have been introduced. The authors focused on small-scale subsistence agriculture and investigated the role of nature and culture in alleviating poverty \citep{Lade}, the role of assets, water, and nutrients \citep{Radosavljevic2020}, the impact of cross-level interactions between individual and community levels \citep{Radosavljevic2021}, the emergence of cross-level poverty traps in agricultural innovation systems \citep{Sanga2024}, or the impact of disease and poor health on persistent poverty \citep{Ngonghala2014,Ngonghala2017}.

Most of poverty trap models are formalized as systems of ordinary differential equations and represent intertwined social-ecological processes observed in real systems. The model in this paper is based on the same principles and is analyzed using the same mathematical methods. The causal structure of the model is based on stylized facts from the published literature and first-hand the experience of the coauthors to specifically represent the SSA system. Combined analytical and numerical techniques allow for the study of the short- and long-term dynamics of an SSA system in a fairly transparent way. Poverty traps are identified with unwanted attractors, and well-being with wanted attractors. 

The number and type of equilibrium points, the size and shape of basins of attraction, bifurcations, and tipping points can be clearly identified, giving a qualitative understanding of the system. The implications for research and management can be significant, but the clarity that models provide is based on simplification of model assumptions and abstraction of system properties. Consequently, regime shifts, resilience of poverty traps, vulnerability to shocks, or other dynamic patterns described in the model may bear little relation to that of the real system. Therefore, dynamical system modeling is best used jointly with empirical research and more complex modeling approaches through an iterative process where model assumptions and results can be tested and validated \citep{Eppinga2024, Radosavljevic2024}.

It is important to emphasize that the objective of dynamical system models, including the model \eqref{model}, is not to accurately represent the dynamics of a particular SSA system, but to explore the qualitative behavior of a stylized system based on empirical and theoretical knowledge. Model analysis does not lead to quantitative predictions or management prescriptions, but it provides a structural understanding, that is, understanding how the structure of the system brings about its behavior \citep{Radosavljevic2023}. Social-ecological dynamics and possible causal relationships are revealed through the manipulation of interactions between state variables and the observation of their consequences \citep{Schluter2024}. Structural understanding is therefore particularly valuable for assessing the effectiveness of interventions, consequences of shocks, and managing transformation toward sustainable outcomes.

\subsection{Implication for sustainable intensification}
Transforming aquaculture in developing countries from extensive to intensive is done with the clear aim of improving livelihoods and food security. However, it entails a set of challenges that are neither fully understood nor easy to assess. The challenges are connected to tension between sustainability and productivity and solving them requires a clearer understanding how intertwined ecological, technological, economic and social processes shape dynamics in multidimensional settings. In agricultural poverty alleviation, for example, there is a tendency to use blanket solutions without paying enough attention to the local context and the social-ecological complexities of the system \citep{Barrett2013, Haider2018}. The consequences of such practices can be dire and even reinforce the dynamics they were set to break \citep{Lade}. The model analysis in Sections 3 and 4 highlights why the local context should be considered in the design of management strategies, how the heterogeneity of SSA producers contributes to the complexity of the system, and how to promote sustainable intensification of SSA systems.

\subsubsection{Effects of the local context} Stability analysis shows that the qualitative behavior of the model (\ref{model}) is highly context dependent, which means that different parameters in the model describe different scenarios in real systems. Depending on the values of the parameters, that is, the local context, the system falls into one of the three different scenarios: a global poverty trap (that is, a monostable assets-free system that exists if conditions \eqref{E1_stab} of \eqref{E2_stab} are fulfilled), a bistable system with poverty trap and well-being attractors (Figure \ref{fig:twoEP}), and a three-stable system with a single poverty trap and two well-being attractors (Figure \ref{fig:threeEP}). 

Model analysis can uncover what an intervention needs to achieve in order to be effective. The global poverty trap can be alleviated only by a series of interventions, where the first needs to transform the system and create a well-being attractor. Consecutive interventions should focus on adjusting initial conditions and placing them in the well-being basin of attraction. In the case of bistable and three-stable systems, it is possible to escape poverty without transforming the system, in which case interventions must place the system in the well-being basin of attraction. Stability analysis, such as the one in Figures \ref{fig:twoEP} and \ref{fig:threeEP}, shows which state variables an intervention should target and how to push the system in the well-being basin of attraction. In Figure \ref{fig:twoEP}A, an intervention should aim to increase assets and/or fish biomass, while in Figure \ref{fig:twoEP}B, it should also decrease nutrients. 

Another alternative for intervention in multistable systems is to increase the well-being basin of attraction to include fixed initial conditions. Bifurcation analysis, such as that in Figures \ref{fig:Model3D_b} and \ref{fig:Model3D_g}, shows what mechanism should be adapted to increase the resilience of the well-being attractor and the size of its basin of attraction. In other words, the analysis shows how changing the local context can open up possibilities for sustainable transformation.

Finally, the analysis reveals that the shape of the basins of attraction depends on the local context.  
For example, the difference between Figures \ref{fig:twoEP}A and \ref{fig:twoEP}B is caused by different parameter values, which symbolize different properties of the real system. Figure \ref{fig:twoEP}A describes the scenario in which it is easier to reach a higher savings rate and the input of runoff nutrients is lower compared to the scenario in Figure \ref{fig:twoEP}B. The same initial conditions (producer wealth, fish biomass, and nutrient level) that lead to well-being in the first scenario could lead to the poverty trap in the second scenario. Thus, for designing efficient interventions, it is crucial to know the number of attractors and what multistability entails.

\subsubsection{Consequences of producers' heterogeneity} Heterogeneity with respect to SSA producers' wealth, used technology, economic conditions, and ecological properties of SSA systems and nearby agricultural lands can determine the success of an intervention. The heterogeneity of the real SSA producers is represented by different initial conditions in the model. Placing initial conditions in different basins of attraction means that their trajectories will converge toward different outcomes. Stability analysis, such as the one in Figures \ref{fig:twoEP} and \ref{fig:threeEP}, can give information on who is more resilient or more at risk, or who in a heterogeneous group of SSA producers would benefit from an intervention and who would be left behind. 

The bifurcation diagrams in Figures \ref{fig:Model3D_b} and \ref{fig:Model3D_g} recap the conclusions of Section 4.1.2, but also show what might be the consequences of increased productivity. Intensification in SSA systems can lead to the emergence of the third stable state, $E_{**}$, which is less desirable than $E_*$ because it has lower values of assets and fish biomass. In the process of intensification by increasing productivity $b$ or the nutrient input $g$, the resilience of $E_*$ to financial shocks is reduced. Trajectories that would converge to $E_*$ in a less intensified system could converge to $E_{**}$ in more intensified systems. The unintended consequences of intensification would be a decrease in producers' income and food security. Similar conclusions hold for changes in the input rate of nutrients, $g$. As expected, a minimal amount of nutrients is necessary for fish growth, and too much nutrients can lead to collapse due to eutrophication and pollution. Given an intermediate amount of nutrients, the system can be bistable or three-stable. The three-stable system could be more harmful than beneficial for poorer producers who risk getting caught in the suboptimal $E_{**}$ state.

\subsection{Links to case studies and implications for management}

Several recent empirical studies have examined the sustainability outcomes of interventions in SSA, highlighting their potential to improve food availability \citep{Wang2024} and enhance livelihoods \citep{DamLam}, although it is not always clear through which pathways diversification stabilize food security and livelihoods. The literature also suggests that the impact of such interventions on poverty remains ambiguous \citep{Belton2011}. \citet{Belton2011} claim that: ``As total volume and value of output are likely to correspond closely to the area of pond under culture, even where poorer producers can be engaged the absolute benefits they derive are likely to be smaller than those of better-off project participants with larger land-holdings". \citet{Cramb2004} supports these findings and states that the impact of SSA is likely to be highly class-differentiated.

The above results can be interpreted with the help of our model. The interventions described in \citet{DamLam} and \citet{Wang2024} increase productivity, likely by increasing parameter $b$ and $g$, and create bistable systems, as shown in the bifurcation diagram in Figures \ref{fig:Model3D_b} and \ref{fig:Model3D_g}. The diagrams show that qualitatively similar results can be obtained in different ways, which reflects the diverse pathways that have been observed in the case studies. Getting more details on particular cases could push the modeling even further. It could help improve the model structure and parameter range and result in a more detailed picture of pathways that lead to sustainable outcomes. 

The bifurcation diagrams in Figures \ref{fig:Model3D_b} and \ref{fig:Model3D_g} can also be used to explore the results of \citet{Belton2011} and \citet{Cramb2004}. According to the model, intensification can create a suboptimal attractor in a three-stable system. The consequence of intensification is that its positive effects could be unevenly distributed among poor and wealthy producers. Wealthy producers are in a better position to benefit from interventions because they are likely in the well-being basin of attraction, while poor producers have a higher chance of being caught in the suboptimal state near the suboptimal attractor $E_{**}$.
Again, having close collaboration between modelers, practitioners and stakeholders could facilitate model development and ensure that the models are empirically grounded. Further stability and bifurcation analyses could be used to explore the intricacies of particular cases, design and assess interventions, and uncover their eventual unintended consequences. 

\subsection{Implications for research}

There are many ways in which the model presented in this paper can be expanded to explore relevant research questions. We list several of them that align with our research interests and open research questions, but the list is in no way complete. 

The dynamics of small-scale aquaculture ponds is affected by temperature variation and is strongly dependent on the biophysical properties of the pond \citep{Bieg, Jobling, Lu}. Including spatio-temporal variability in models would increase their mathematical complexity because it requires nonautonomous systems of ordinary or partial differential equations. However, it would allow us to explore the interplay between spatio-temporal patterns with social-ecological processes in the system. Including more details on the ecological side could help explore polyculture ponds \citep{Milstein} and the effects of disease spread on poverty \citep{Hoover}.

The problems in governing SSA systems can be seen as social dilemmas in which shared water or space are examples of common pool resources \citep{Partelow2022}. Demographic heterogeneity, that is, differences in SSA producers' wealth, training, opinions, and perceived risks, could be reflected in their preferred strategies and decisions and play an important role in shaping the SSA dynamics \citep{Rahman, Nagel2024}. The effects of social norms and management strategies on producers' decisions, and in turn on SSA dynamics, can be explored using combinations of dynamical systems, evolutionary game-theoretic and agent-based models. 

There is a great deal of uncertainty and lack of firm knowledge on financial, climatic, and environmental shocks and their effects on aquaculture dynamics \citep{Luna}. Dynamical systems typically focus on asymptotic behavior and processes that last forever, but including transient analysis can give answers to questions concerning shorter time periods and instantaneous processes. These results could contribute to understanding out-of-equilibrium dynamics and inform stakeholders and management about efficient ways to adapt and respond to shocks.


\section*{CRediT authorship contribution statement}
SR- Conceptualization, Methodology, Software,
Formal analysis, Writing - original draft, Funding acquisition. EV - Conceptualization, Formal analysis, Writing - original draft, Funding acquisition. FA - Formal analysis, Writing - original draft, Funding acquisition. QW - Conceptualization, Writing - original draft, Funding acquisition. JS - Conceptualization, Writing - original draft, Funding acquisition. AG - Conceptualization, Writing - original draft, Funding acquisition.

\section*{Declaration of Competing Interest}
The authors declare that they have no known competing financial interests or personal relationships that could have appeared to influence the work reported in this paper.

\section*{Acknowledgments}
SR has been supported by the Swedish Research Council FORMAS (grant number 2021-01840).
EV and FA have been partially supported by the project
``Analisi di metodi ed algoritmi in teoria dell'approssimazione e modelli in biologia ed ecologia''
(Analysis of methods and algorithms in approximation theory and models in biology and ecology) of the Department of
Mathematics ‘‘Giuseppe Peano’’ of the University of Turin. EV and FA are members of the Italian research group ``Gruppo Nazionale per il Calcolo Scientifico" (GNCS) of the ``Istituto Nazionale di Alta Matematica" (INdAM). AG, QW, and JS acknowledge the financial support of WorldFish under the technical support grant "Characteristics and impacts of aquatic food systems".


\section*{Appendix}

In this appendix, we provide mathematical details related to the feasibility of the assets-free equilibrium, $E_2$, and coexistence, $E_*$, summarized in Table \ref{tab_eq} of Section \ref{analysis}.

\subsection*{Feasibility of assets-free point, $E_2$}

From the second equilibrium equation of the model \eqref{modelSimple}, i.e., equilibrium equation for $F$, we obtain $F=\Psi(N)$, where
\begin{equation}\label{Psi}
\Psi(N)= \frac {ruN} {c(v+N^2)} - \frac {m+h}c.
\end{equation}

On the other hand, from the third equilibrium equation, i.e., equilibrium equation for $N$, we explicitly get $F= \Theta (N)$, where
\begin{equation*}
\Theta(N)= \frac {v+N^2} {uN} (k - \ell N).
\end{equation*}

Thus, for the feasibility of $E_2$ we can look for sufficient conditions to have at least one intersection point between the curves $\Psi$ and $\Theta$ in the first quadrant of the $N$-$F$ plane.

\subsubsection*{Study of the curve $\Psi$}

The function $\Psi(N)$ for $N\ge 0$ is a gamma-like function. Its intersection with the vertical axis and its horizontal asymptote are, respectively, given by
$\left( 0, - \frac {m+h}c \right)$ and $F_a = - \frac {m+h}c$.
Therefore, it lies outside the feasible region unless its peak, located at $N=N_*$,
has a positive height, $\Psi (N_*) > 0$.

The maximum can easily be established by differentiation, giving
\begin{equation}\label{N*}
N_*= \sqrt {v} \quad \text{and} \quad 
\Psi(N_*)= \frac {ru \sqrt{v}} {2cv} - \frac {m+h}c,
\end{equation}
and the feasibility condition
\begin{equation}\label{Psi_feas}
ru > 2 (m+h) \sqrt{v}.
\end{equation}

Then, the function $\Psi(N)$ intersects the horizontal axis whenever
$$
(v+N^2)(m+h)-ruN = 0.
$$
The latter is a quadratic
equation, whose roots are explicitly found:
\begin{equation*}
N_{\pm} = \frac 1{2(m+h)} \left( ru \pm \sqrt {r^2u^2  - 4 v (m+h)} \right).
\end{equation*}
Clearly $N_{\pm} \ge 0$ and $\Psi(N)$ is nonnegative in $[N_-,N_+]$, with $0<N_-< N_* <N_+$.

\subsubsection*{Study of the curve $\Theta$}

The function $\Theta(N)$ has a vertical asymptote on the coordinate axis $N=0$ and is positive for $0<N< N_0$,
where $N_0= \frac k{\ell}$.

We also find
\begin{equation*}
\frac {d\Theta}{dN}= u \frac {N^2 -v } {u^2N^2} (k - \ell N)
- \ell \frac {v+N^2} {uN}.
\end{equation*}
Therefore, it is not immediately clear whether the function is monotonically decreasing. The positive condition of the first derivative 
is equivalent to the cubic inequality
$$
- 2 \ell N^3 + k N^2 - vk >0,
$$
so that by the Descartes rule there are two positive roots. The question is whether they are located
in the interval $[0, N_0]$. In case they are not, in the very same interval $\Theta (N)$ is monotonically
decreasing and no multiple intersections with the other function arising from the equilibrium equation of $N$ would be possible. Conversely, $\Theta$ exhibits a kink in the feasible region and multiple intersections could arise.

The second derivative of the function is
\begin{equation*}
\frac {d^2\Theta}{dN^2}= - 6 \ell N^2 + k.
\end{equation*}
It is nonnegative for $0<N<\widehat N$, where
$\widehat N= \sqrt {\frac k{6\ell}}$. 
Therefore, we find $N_0 < \widehat N$ for 
\begin{equation}\label{Theta_deriv2_positive}
6 k < \ell,
\end{equation}
ensuring the monotonicity of $\Theta(N)$ in $[0, N_0]$.

We can further investigate the condition $\Theta '(N)<0$ in $[0, N_0]$. It is equivalent to
\begin{equation}\label{L-Rtilde}
L (N) = \ell \frac {v+N^2}{uN} < \frac {N^2-v}{uN^2} (k-\ell N) = \widetilde R (N)
= \frac 1{uN^2} R (N) .
\end{equation}
Now, $L(N)\ge 0$ for every $N \in [0, N_0]$. On the other hand, $R (N)\ge 0$ for $N \in [N_*, N_0]$.
This means that $L(N)<\widetilde R(N)$ for $0<N<\sqrt{v},$ $N> \frac k{\ell}$.
In $N \in [N_*, N_0]$ we need to investigate the conditions for which $L$ and $\widetilde R$ intersect,
as they are both positive.
Rewriting (\ref{L-Rtilde}) in the simplified form
$$
\ell N (v+N^2) < (N^2-v) (k-\ell N),
$$
we can finally establish the condition for the intersections of $N L(N)$ and $R(N)$ for $N \in [N_*, N_0]$.
Differentiating, we find
$$
\frac {dR}{dN} = -3N^2 + kN + \ell v,
$$
from which the maximum is attained at the point
$$
N_{**}^+ = \frac 16 \left[ k + \sqrt {k^2+12 \ell v} \right]
$$
and the condition that must be satisfied is
\begin{equation}\label{N**}
N_{**}^+ L(N_{**}^+) > R(N_{**}^+) .
\end{equation}
Summarizing, (\ref{N**}) ensures that (\ref{L-Rtilde}) holds, which means that $\Theta(N)$ is
monotonically decreasing in $[0,N_0]$.

\subsubsection*{Intersections of the curves $\Psi$ and $\Theta$}

Assume now the monotonicity of $\Theta(N)$, that is, (\ref{Theta_deriv2_positive}) or (\ref{N**}).
We need to find the intersections of the curves $\Psi$ and $\Theta$, as these provide the assets-free equilibrium points, in the first quadrant of the $N$-$P$ plane.

There are four possible situations, the first two of which do not lead to any feasible
intersection, namely 
\begin{equation}\label{E2_no}
 N_0<N_- \quad \text{and} \quad N_0>N_+, 
\end{equation}
conditions that will therefore be disregarded. We focus instead on the following two:
\begin{eqnarray}\label{E2!}
N_-<N_0<N_+, \\ \label{E2-saddle}
N_*<N_+ <N_0.
\end{eqnarray}

The condition (\ref{E2!}) ensures the uniqueness of the intersection, therefore, giving an
equilibrium point $E_2$. 

On the other hand, (\ref{E2-saddle}) does not always ensure the existence
of the intersection; in the case where $\Psi$ and $\Theta$ are tangent to each other uniqueness is guaranteed.
In fact, note that $\Psi$ is concave and $\Theta$ is convex. However,
if they intersect, there will be a pair of equilibria $E_2$ through a saddle-node bifurcation.
For this to occur, we need also $\Psi (N_*) < \Theta (N_*)$, which using (\ref{N*}) can explicitly
be written as
\begin{equation}\label{saddle-condition}
2 \frac vu (k- \ell \sqrt {v}) < \frac {ru}{2c} - \sqrt {v} \frac {m+h}c.
\end{equation}

\subsection*{Feasibility of coexistence, $E_*$}

The second equilibrium equation, that is, the equilibrium equation for $F$, gives the function $F = \Psi(N)$, with (\ref{Psi}), already
investigated for the assets-free equilibrium, $E_2$. However, the difference is that in the previous subsection this was a curve in the $N-F$ plane, while here, in the three-dimensional space, it is a cylinder
with the axis parallel to the $A$ axis.

Then, from the first equilibrium equation, i.e., the equilibrium equation for $A$, solving for $F$, we find the surface
\begin{eqnarray}\label{chi}
F = \chi (A) = \frac q{bA^2} (p+A^2).
\end{eqnarray}

Finally, from the last equilibrium equation for $N$ we get another surface, namely
\begin{eqnarray}\label{Phi}
F = \Phi (N,A) = \frac {(k-\ell N) (v+N^2) (z+A)} {uN(z+A) - gA (v+N^2)}.
\end{eqnarray}

In this case, for the coexistence feasibility, we can look for sufficient conditions to have at least one intersection point of these three surfaces, $ \Psi$, $\chi$, and $\Phi$, in the first octant of the $A$-$N$-$F$ space.

\subsubsection*{Study of the surface $\chi$}

The function $\chi (A)$, being independent of $N$, represents a cylinder with the axis parallel to the $N$ axis.
Its intersection with the $N=0$ coordinate plane is a hyperbola-like function, with a vertical
asymptote on the $F$ axis and a horizontal one located at
\begin{eqnarray}\label{F_infty}
F_{\infty} = \frac qb.
\end{eqnarray}

This function is monotonically decreasing in view of the fact that
$$
\frac {d \chi}{dA} = - \frac 2{bA^2}<0.
$$

\subsubsection*{Study of the surface $\Phi$}

Let us define the following quantity, the denominator of (\ref{Phi})
\begin{equation*}
D_{\Phi} = uzN + uNA - gv A - g A N^2 = - N_{\mathcal{F}} (N) - A D_{\mathcal{F}} (N),
\end{equation*}
where
\begin{equation*}
N_{\mathcal{F}} (N) = uz N \quad \text{and} \quad D_{\mathcal{F}} (N) = g N^2 - uN + gv.
\end{equation*}
To assess the regions for which $\Phi$ is feasible, we need to study the sign of its denominator, $D_{\Phi}$, since its numerator, $N_{\Phi}$, is easily seen to be positive for
\begin{equation}\label{N_Phi_pos}
N < N_0.
\end{equation}

Now, $D_{\Phi}>0$ is equivalent to $A D_{\mathcal{F}} < N_{\mathcal{F}}$. This inequality holds trivially
for $D_{\mathcal{F}}<0$.
Conversely, for $D_{\mathcal{F}} > 0$ it reduces to
$$
A \le {\mathcal{F}} (N) = \frac {N_{\mathcal{F}}}{D_{\mathcal{F}}}.
$$
The roots of $D_{\mathcal{F}}=0$ are
$$
N_{u,\ell} = \frac 1{2g} \left[ u \pm \sqrt{u^2 - 4 g^2 v} \right],
$$
where the subscript $\ell$ (lower) corresponding to the minus sign and $u$ (upper) to the plus sign.
Thus, in this case $D_{\mathcal{F}}>0$ for $N<N_{\ell}$ and $N>N_u$.
Combining these results, we find
$D_{\Phi} > 0$ for the following alternative cases:
\begin{eqnarray}\label{D_Phi_pos}
D_{\mathcal{F}}>0:  A\le {\mathcal{F}}; \qquad D_{\mathcal{F}}>0:  \mbox{always true}.
\end{eqnarray}
The function $A={\mathcal{F}}$
has two branches in the first quadrant. The left one crosses the origin and raises up to a
vertical asymptote located at $N=N_{\ell}$. In $(N_{\ell},N_u)$ the function is negative,
while it decreases from another vertical asymptote at $N=N_u$ to approach the horizontal axis
for $N\rightarrow +\infty$.

We now concentrate on the feasibility of the function $\Phi(N,A)$.
On $A={\mathcal{F}}$, the surface $\Phi$ has a vertical asymptote.
On the other hand, on $N=N_0$ it vanishes.
The above two lines, $A={\mathcal{F}}$ and $N=N_0$, intersect at the point
\begin{equation}\label{A0}
A_0= {\mathcal{F}}(N_0) = \frac {\ell kuz}{g\ell ^2 v- k\ell u + gk^2},
\end{equation}
feasible if
$$
g\ell ^2 v + gk^2 > k\ell .
$$
Note that at the point $A_0$ the surface $\Phi$ does not have a limit, because if the point
is approached along the curve ${\mathcal{F}}$, the surface grows without limit, while if $A_0$
is approached along the line $N=N_0$ the surface vanishes.

We have to distinguish two different
alternative situations leading to $\Phi(N,A)>0$:
\begin{equation}\label{a}
N_{\Phi} > 0, \quad D_{\Phi} > 0
\end{equation}
or
\begin{equation}\label{b}
N_{\Phi} < 0, \quad D_{\Phi} < 0.
\end{equation}

In the case (\ref{a}), we need $N<N_0$, with no other conditions, if $D_{\mathcal{F}}<0$. Instead, we need
$N<N_0$ and $A\le {\mathcal{F}}(N)$ for $D_{\mathcal{F}}>0$, compare (\ref{N_Phi_pos}) and (\ref{D_Phi_pos}).
Geometrically, the latter means
that the feasible region in the $N$-$A$ plane lies below the two positive branches
of $A={\mathcal{F}}(N)$ and includes also the half stripe in the first quadrant bounded
below by the interval $(N_{\ell},N_u)$. We must further distinguish three subcases
depending on the location of $N_0$ with respect to $(N_{\ell},N_u)$:
\begin{itemize}
\item[(a1)] $N_0 < N_{\ell} < N_u$: the surface $\Phi$ is positive in the ``triangular''
region $\Omega_{a1}$, with a vertex at the point $(N_0,A_0)$,
bounded above by the left branch of ${\mathcal{F}}$, on the right by the vertical line $N=N_0$
and below by the coordinate axis $N$;
\item[(a2)] $N_{\ell} < N_0 < N_u$: the surface $\Phi$ is positive in the region $\Omega_{a2}$ bounded below by the coordinate axis $N$ and above by the left branch of ${\mathcal{F}}$ for $N<N_{\ell}$, and in the half-stripe for $N_{\ell} < N < N_0$;
\item[(a3)] $N_{\ell} < N_u < N_0$: the surface $\Phi$ is positive in the region $\Omega_{a3}$ bounded below by the coordinate axis $N$,
bounded above by the left branch of ${\mathcal{F}}$ for $N<N_{\ell}$, in the half-stripe 
for $N_{\ell} < N < N_u$ and bounded above by
the right branch of ${\mathcal{F}}$ for $N_u<N<N_0$.
\end{itemize}

In the case (\ref{b}), we need 
$N>N_0$ and $A > {\mathcal{F}}(N)$ for $D_{\mathcal{F}}>0$; the condition does not hold if $D_{\mathcal{F}}<0$.
Hence, $\Phi(N,A)$ is positive only above the function ${\mathcal{F}}$ whenever this is positive.
Here, too, there are three subcases:
\begin{itemize}
\item[(b1)] $N_0 < N_{\ell} < N_u$: the surface $\Phi$ is positive in the 
region $\Omega_{b1}$ 
above the left branch of ${\mathcal{F}}$ for $N_0<N<N_{\ell}$ and
above the right branch of ${\mathcal{F}}$ for $N>N_u$;
\item[(b2)] $N_{\ell} < N_0 < N_u$: the surface $\Phi$ is positive 
in the region $\Omega_{b2}$ 
above the right branch of ${\mathcal{F}}$, i.e., for $N > N_u$;
\item[(b3)] $N_{\ell} < N_u < N_0$: the surface $\Phi$ is positive 
in the region $\Omega_{b3}$ 
above the right branch of ${\mathcal{F}}$ for $N>N_0$.
\end{itemize}

Note also that the surface $\Phi$ in the regions that are unbounded has different behaviors, namely
\begin{equation*}
\lim_{N\rightarrow +\infty} \Phi(N,A) = + \infty,
\end{equation*}
while
\begin{equation*}
\lim_{A\rightarrow +\infty} \Phi(N,A) = 
\frac {(k-\ell N) (v+N^2)} {uN - g (v+N^2)},
\end{equation*}
whose value depends on $N$ but it is finite.

\subsubsection*{Study of the curve $\widetilde\Lambda = \chi \cap \Psi$}

In the following analysis, some cases will hinge on the mutual behavior of $\Phi$ and $\widetilde\Lambda = \chi \cap \Psi$
as $a\rightarrow + \infty$. The latter is above the surface if the following inequality holds, and conversely:
\begin{equation}\label{comp_Phi_limit_A}
\frac {(k-\ell N) (v+N^2)} {uN - g (v+N^2)} < \frac qb.
\end{equation}

We now turn to studying the curve $\widetilde\Lambda = \chi \cap \Psi$.
Because the former is above the plane $F=qb^{-1}$ and the latter has the
height of the maximum $F=\Psi(N_*,\widehat A)$, $\widehat A$ being an arbitrary value as $\Psi$ is a cylinder,
they can intersect only if
$\Psi(N_*,\widehat A) > qb^{-1}$, a condition that explicitly becomes
\begin{equation}\label{chi_intersect_Psi}
bru > 2cqv + 2b (m+h) v.
\end{equation}

Because $\chi$ raises up to infinity for $A=0$, i.e., on the $N$-$F$ coordinate plane,
for increasing $A$ it decreases toward its horizontal asymptote. The first intersection with the cylinder
$\Psi$ must occur at a point $X=(A_X,F_X,N_X)$, with $N_X=N_*=\sqrt {v} >0$.
We must then have $\chi(N_*,A_X) = \Psi(N_*,A_X)$, from which
follows
$$
A_X = \sqrt {\frac {2cpqv} {bru-2cqv - 2 bv(m+h)}}.
$$
The intersection exists only if $A_X\ge 0$, that is, if (\ref{chi_intersect_Psi}) holds.
Finally, we explicitly have
$$
X = (A_X, \Psi(\sqrt{v}),\sqrt{v}).
$$

The curve $\widetilde\Lambda = \chi \cap \Psi$ originates from $X$ and consists of two branches,
$\widetilde\Lambda_-$ and $\widetilde\Lambda_+$, respectively, for $N\le N_*$ and $N\ge N_*$.
In view of the fact that they lie on $\chi$, as $A\rightarrow + \infty$ both
approach $F=qb^{-1}$ and on this plane also, respectively, approach the values $N=N_{\mp}^{\infty}$, with
$N_-<N_-^{\infty}$ and $N_+^{\infty}<N_+$.
The latter is obtained by imposing that $\Psi(N)=qb^{-1}$. We find
\begin{equation*}
N_{\mp}^{\infty} = \frac 1{2[cq+b(m+h)]} \left[ bru \pm \sqrt{ b^2r^2u^2 - 4[cq+b(m+h)]^2 v}\right].
\end{equation*}

\subsubsection*{Intersections of the curve $\widetilde\Lambda$ with the surface $\Phi$}

Coexistence is obtained from the intersection of the curve $\widetilde\Lambda$ with the surface $\Phi$.
Several situations can arise, due to the various cases (a1)-(a3) and (b1)-(b3) examined
above, in combination with the location of the point $X$, the branches of $\widetilde\Lambda$, and their
asymptotes at $N_{\mp}^{\infty}$.
The existence of the intersection relies on the fact that in the phase space the curve $\widetilde\Lambda$
approaches the horizontal plane
$F=F_{\infty}$, see (\ref{F_infty}), and that on ${\mathcal{F}}(N)$ the function $\Phi$ has a vertical asymptote.

Our discussion focuses mainly on the location of the projection of the point $X$, $(N_X,A_X)$
on the $N$-$A$ coordinate plane, from which the projection $A=\Lambda(N)$
of the curve $\widetilde \Lambda$ originates, and 
the feasible regions where $\Phi \ge 0$ discovered in (a1)-(a3) and (b1)-(b3) above.
Note that the curve $\Lambda$ has vertical asymptotes at $N=N_{\mp}^{\infty}$.

The cases that can arise are many, too many to list exhaustively. In addition, in several of them it is not clear whether the intersection is unique or in some cases multiple (in general double). In the latter case, there would most likely be saddle-node bifurcations giving rise to pairs of equilibria, but to specify the conditions under which they arise would be very difficult, also because the coordinates of the coexistence equilibrium point are not explicit. We therefore confine ourselves to list the cases where the existence and uniqueness of the coexistence equilibrium occurs and disregard other more complicated situations. See Tables \ref{tab_a1}, \ref{tab_a2}, \ref{tab_a3}, \ref{tab_b1}, \ref{tab_b2}, and \ref{tab_b3}.

\begin{table}[ht]
\footnotesize
\begin{tabularx}{\textwidth}{p{5cm}  p{10cm}}\\
\toprule
\bf{Intersections} & \bf{\hspace*{1cm} Conditions} \\ 
\toprule
no intersection & \hspace*{1cm} if $N_0 < N_-^{\infty}$ \\
\midrule
unique intersection & \begin{itemize}
\vspace*{-0.25cm}
    \item for $N_-^{\infty} < N_0 < N_*$, iff $\Lambda(N_0) < A_0$
    \item for $N_*< N_0 < N_{\ell} < N_+^{\infty}$, iff $\Lambda(N_0) < A_0$ 
   \vspace*{-0.25cm}
\end{itemize} \\
\midrule
two intersections & \hspace*{1cm} for $N_* < N_+^{\infty} < N_0 < N_{\ell}$, if $\widetilde \Lambda (N_*,A_X) > \Phi (N_*,A_X)$ \\
\bottomrule
\end{tabularx}
\smallskip
\caption{Intersections of $\widetilde\Lambda$ with $\Phi$ in the case (a1).}
\label{tab_a1}
\end{table}

\newpage

\begin{table}[ht]
\footnotesize
\begin{tabularx}{\textwidth}{p{5cm}  p{10cm}}\\
\toprule
\bf{Intersections} & \bf{\hspace*{1cm} Conditions} \\ 
\toprule
no intersection & \begin{itemize}
\vspace*{-0.25cm}
    \item if $N_0 < N_-^{\infty}$
    \item for $N_+^{\infty} < N_{\ell}$,
if $\widetilde \Lambda (N_*,A_X) > \Phi (N_*,A_X)$
   \vspace*{-0.25cm}
\end{itemize}\\
\midrule
unique intersection in $[N_-^{\infty}, N_0]$ & 
\begin{itemize}
\vspace*{-0.25cm}
    \item for $N_{\ell} < N_-^{\infty} < N_0 < N_+^{\infty}$,
 if not (\ref{comp_Phi_limit_A})
 \item for $N_-^{\infty} < N_{\ell} < N_0 < N_+^{\infty}$
 \vspace*{-0.25cm}
\end{itemize} \\
\midrule
two intersections in $[N_-^{\infty}, N_+^{\infty}]$ & 
\begin{itemize}
\vspace*{-0.25cm}
    \item for $N_{\ell} < N_-^{\infty} < N_+^{\infty} < N_0$
    \begin{itemize}
    \item if $\widetilde \Lambda (N_*,A_X) > \Phi (N_*,A_X)$ and not (\ref{comp_Phi_limit_A}) 
    \item if $\widetilde \Lambda (N_*,A_X) < \Phi (N_*,A_X)$ and (\ref{comp_Phi_limit_A})
    \end{itemize}
\item for $N_+^{\infty} < N_{\ell}$, if $\widetilde \Lambda (N_*,A_X) < \Phi (N_*,A_X)$
\item for $N_-^{\infty} < N_{\ell} < N_+^{\infty} < N_0$
\begin{itemize}
    \item if $\widetilde \Lambda (N_*,A_X) > \Phi (N_*,A_X)$ and not (\ref{comp_Phi_limit_A})
    \item if $\widetilde \Lambda (N_*,A_X) < \Phi (N_*,A_X)$ and (\ref{comp_Phi_limit_A})
\end{itemize}
    \vspace*{-0.25cm}
\end{itemize} \\
\bottomrule
\end{tabularx}
\smallskip
\caption{Intersections of $\widetilde\Lambda$ with $\Phi$ in the case (a2).}
\label{tab_a2}
\end{table}

\begin{table}[ht]
\footnotesize
\begin{tabularx}{\textwidth}{p{5cm}  p{10cm}}\\
\toprule
\bf{Intersections} & \bf{\hspace*{1cm} Conditions} \\ 
\toprule
no intersection & \begin{itemize}
\vspace*{-0.25cm}
    \item if $N_0 < N_-^{\infty}$ 
    \item for $N_u < N_-^{\infty} < N_0 < N_+^{\infty}$,
if $\Lambda (N_0) > {\mathcal{F}} (N_0)$
\item for $N_{\ell} < N_-^{\infty} < N_u < N_0 < N_+^{\infty}$, 
if $\Lambda (N_0) < {\mathcal{F}} (N_0)$
\item for $N_u < N_-^{\infty} < N_+^{\infty} < N_0$, if $\Lambda (N_*) > {\mathcal{F}} (N_*)$
\item for $N_-^{\infty} < N_+^{\infty} < N_{\ell} < N_u < N_0$,
if $\Lambda (N_X) > {\mathcal{F}} (N_X)$
    \vspace*{-0.25cm}
\end{itemize}
\\
\midrule
unique intersection in $[N_-^{\infty}, N_0]$ & \begin{itemize}
\vspace*{-0.25cm}
    \item for $N_u < N_-^{\infty} < N_0 < N_+^{\infty}$, 
if $\Lambda (N_0) < {\mathcal{F}} (N_0)$
\item for $N_{\ell} < N_-^{\infty} < N_u < N_0 < N_+^{\infty}$, 
if $\Lambda (N_0) > {\mathcal{F}} (N_0)$
    \vspace*{-0.25cm}
\end{itemize} \\
\midrule
unique intersection 
in $[N_-^{\infty}, N_u]$ & \hspace*{1cm} for $N_-^{\infty} < N_{\ell} < N_u < N_0 < N_+^{\infty}$, if $\Lambda (N_0) < {\mathcal{F}} (N_0)$\\
\midrule
two intersections in $[N_-^{\infty}, N_+^{\infty}]$ & \begin{itemize}
\vspace*{-0.25cm}
    \item for $N_-^{\infty} < N_+^{\infty} < N_{\ell} < N_u < N_0$, if $\Lambda (N_X) < {\mathcal{F}} (N_X)$
    \item for $N_{\ell} < N_-^{\infty} < N_+^{\infty} < N_u < N_0$
    \begin{itemize}
    \item if $\widetilde\Lambda (N_X,A_X) < \Phi (N_X,A_X)$ and not (\ref{comp_Phi_limit_A})
    \item if $\widetilde\Lambda (N_X,A_X) > \Phi (N_X,A_X)$ and (\ref{comp_Phi_limit_A})
    \end{itemize}
     \item for $N_{\ell} < N_-^{\infty} < N_u < N_+^{\infty} < N_0$
     \begin{itemize}
    \item if $\widetilde\Lambda (N_X,A_X) < \Phi (N_X,A_X)$ and not (\ref{comp_Phi_limit_A}) 
    \item if $\widetilde\Lambda (N_X,A_X) > \Phi (N_X,A_X)$ and (\ref{comp_Phi_limit_A})
    \end{itemize}
\item for $N_u < N_-^{\infty} < N_+^{\infty} < N_0$, if $\Lambda (N_*) < {\mathcal{F}} (N_*)$
\item for $N_-^{\infty} < N_{\ell} < N_u < N_+^{\infty} < N_0$
\item for $N_-^{\infty} < N_{\ell} < N_u < N_0 < N_+^{\infty}$,
if $\Lambda (N_0) > {\mathcal{F}} (N_0)$
\item for $N_-^{\infty} < N_{\ell} < N_+^{\infty} < N_u < N_0$
\begin{itemize}
    \item if $\widetilde\Lambda (N_X,A_X) < \Phi (N_X,A_X)$ and not (\ref{comp_Phi_limit_A}) 
    \item if $\widetilde\Lambda (N_X,A_X) > \Phi (N_X,A_X)$ and (\ref{comp_Phi_limit_A})
\end{itemize}
 \vspace*{-0.25cm}
\end{itemize} \\
\bottomrule
\end{tabularx}
\smallskip
\caption{Intersections of $\widetilde\Lambda$ with $\Phi$ in the case (a3).}
\label{tab_a3}
\end{table}

\newpage

\begin{table}[ht]
\footnotesize
\begin{tabularx}{\textwidth}{p{5cm}  p{10cm}}\\
\toprule
\bf{Intersections} & \bf{\hspace*{1cm} Conditions} \\ 
\toprule
no intersection & \begin{itemize}
\vspace*{-0.25cm}
    \item for $N_-^{\infty} < N_+^{\infty} < N_0 < N_{\ell} < N_u$
\item for $N_-^{\infty} < N_0 < N_{\ell} < N_+^{\infty} < N_u$, if
$\Lambda (N_X) < {\mathcal{F}} (N_X)$
\item for $N_-^{\infty} < N_0 < N_{\ell} < N_u < N_+^{\infty}$,
if $\Lambda (N_X) < {\mathcal{F}} (N_X)$ and not (\ref{comp_Phi_limit_A}) 
\item for $N_0 < N_-^{\infty} < N_+^{\infty} < N_{\ell} < N_u$,
if not (\ref{comp_Phi_limit_A}) and $\widetilde\Lambda (N_X,A_X) < \Phi (N_X,A_X)$ 
\item for $N_0 < N_{\ell} < N_-^{\infty} < N_+^{\infty} < N_u$
\item for $N_0 < N_{\ell} < N_u < N_-^{\infty} < N_+^{\infty}$,
if not (\ref{comp_Phi_limit_A}) 
   \vspace*{-0.25cm}
\end{itemize} \\
\midrule
one intersection & \begin{itemize}
\vspace*{-0.25cm}
 \item for $N_-^{\infty} < N_0 < N_+^{\infty} < N_{\ell} < N_u$,
if not (\ref{comp_Phi_limit_A})
\item for $N_-^{\infty} < N_0 < N_{\ell} < N_+^{\infty} < N_u$, 
if $\Lambda (N_X) > {\mathcal{F}} (N_X)$
\item for $N_-^{\infty} < N_0 < N_{\ell} < N_u < N_+^{\infty}$
\begin{itemize}
    \item if $\Lambda (N_X) > {\mathcal{F}} (N_X)$ and not (\ref{comp_Phi_limit_A})
    \item if $\Lambda (N_X) < {\mathcal{F}} (N_X)$ and (\ref{comp_Phi_limit_A}) 
\end{itemize}
 \vspace*{-0.25cm}
\end{itemize} \\
\midrule
one intersection in $[N_-^{\infty}, N_{\ell}]$
 & \begin{itemize}
\vspace*{-0.25cm}
\item for $N_0 < N_-^{\infty} < N_{\ell} < N_+^{\infty} < N_u$
\item for $N_0 < N_-^{\infty} < N_{\ell} < N_u < N_+^{\infty}$
 \vspace*{-0.25cm}
\end{itemize} \\
\midrule
one intersection in $[N_u, N_+^{\infty}]$
 & \hspace*{1cm} for $N_0 < N_{\ell} < N_-^{\infty} < N_u < N_+^{\infty}$,
 if (\ref{comp_Phi_limit_A}) \\
\midrule
two intersections & \begin{itemize}
\vspace*{-0.25cm}
 \item for $N_-^{\infty} < N_0 < N_{\ell} < N_u < N_+^{\infty}$,
if $\Lambda (N_X) > {\mathcal{F}} (N_X)$ and (\ref{comp_Phi_limit_A})
\item for $N_0 < N_-^{\infty} < N_+^{\infty} < N_{\ell} < N_u$,
if (\ref{comp_Phi_limit_A}) and $\widetilde\Lambda (N_X,A_X) < \Phi (N_X,A_X)$ 
\item for $N_0 < N_-^{\infty} < N_+^{\infty} < N_{\ell} < N_u$
\begin{itemize}
    \item if $\Lambda (N_X) < {\mathcal{F}} (N_X)$ and (\ref{comp_Phi_limit_A})
    \item if $\Lambda (N_X) > {\mathcal{F}} (N_X)$ and not (\ref{comp_Phi_limit_A})
\end{itemize}
\item for $N_0 < N_{\ell} < N_u < N_-^{\infty} < N_+^{\infty}$, if $\Lambda (N_X) < {\mathcal{F}} (N_X)$ and (\ref{comp_Phi_limit_A})
\item for $N_0 < N_-^{\infty} < N_{\ell} < N_u < N_+^{\infty}$, if (\ref{comp_Phi_limit_A})
   \vspace*{-0.25cm}
\end{itemize} \\
\bottomrule
\end{tabularx}
\smallskip
\caption{Intersections of $\widetilde\Lambda$ with $\Phi$ in the case (b1).}
\label{tab_b1}
\end{table}

\newpage

\begin{table}[ht]
\footnotesize
\begin{tabularx}{\textwidth}{p{4.5cm}  p{10.5cm}}\\
\toprule
\bf{Intersections} & \bf{\hspace*{1cm} Conditions} \\ 
\toprule
no intersection 
 & \begin{itemize}
\vspace*{-0.25cm}
\item for $N_-^{\infty} < N_+^{\infty} < N_{\ell} < N_u$
\begin{itemize}
    \item if $\Lambda (N_X) > {\mathcal{F}} (N_X)$ and (\ref{comp_Phi_limit_A})
    \item if $\Lambda (N_X) < {\mathcal{F}} (N_X)$ and not (\ref{comp_Phi_limit_A})
\end{itemize}
\item for $N_-^{\infty} < N_{\ell} < N_u < N_+^{\infty}$, if not (\ref{comp_Phi_limit_A})
\item for $N_{\ell} < N_-^{\infty} < N_+^{\infty} < N_u$
\item for $N_{\ell} < N_-^{\infty} < N_u < N_+^{\infty}$,
if not (\ref{comp_Phi_limit_A})
\item for $N_{\ell} < N_u < N_-^{\infty} < N_+^{\infty}$
\begin{itemize}
    \item if $\Lambda (N_X) > {\mathcal{F}} (N_X)$ and (\ref{comp_Phi_limit_A}) 
   \item if $\Lambda (N_X) < {\mathcal{F}} (N_X)$ and not (\ref{comp_Phi_limit_A})
\end{itemize}
 \vspace*{-0.25cm}
\end{itemize} \\
\midrule
one intersection
 & \begin{itemize}
\vspace*{-0.25cm}
\item for $N_-^{\infty} < N_+^{\infty} < N_{\ell} < N_u$,
if (\ref{comp_Phi_limit_A})
\item for $N_{\ell} < N_-^{\infty} < N_u < N_+^{\infty}$,
if (\ref{comp_Phi_limit_A}) 
\vspace*{-0.25cm}
\end{itemize} \\
\midrule
two intersections 
 & \begin{itemize}
\vspace*{-0.25cm}
\item for $N_-^{\infty} < N_+^{\infty} < N_{\ell} < N_u$,
if $\Lambda (N_X) > {\mathcal{F}} (N_X)$ and (\ref{comp_Phi_limit_A})
\item for $N_-^{\infty} < N_{\ell} < N_+^{\infty} < N_u$,
if (\ref{comp_Phi_limit_A})
\item for $N_-^{\infty} < N_{\ell} < N_u < N_+^{\infty}$,
if (\ref{comp_Phi_limit_A})
\item for $N_{\ell} < N_u < N_-^{\infty} < N_+^{\infty}$
\begin{itemize}
    \item if $\Lambda (N_X) > {\mathcal{F}} (N_X)$ and not (\ref{comp_Phi_limit_A}) 
    \item if $\Lambda (N_X) < {\mathcal{F}} (N_X)$ and (\ref{comp_Phi_limit_A}) 
\end{itemize}
\vspace*{-0.25cm}
\end{itemize} \\
\bottomrule
\end{tabularx}
\smallskip
\caption{Intersections of $\widetilde\Lambda$ with $\Phi$ in the case (b2).}
\label{tab_b2}
\end{table}

\begin{table}[ht]
\footnotesize
\begin{tabularx}{\textwidth}{p{3.5cm}  p{11.5cm}}\\
\toprule
\bf{Intersections} & \bf{\hspace*{1cm} Conditions} \\ 
\toprule
no intersection & \begin{itemize}
\vspace*{-0.25cm}
\item for $N_-^{\infty} < N_+^{\infty} < N_{\ell} < N_u < N_0$
\item for $N_-^{\infty} < N_{\ell} < N_+^{\infty} < N_u < N_0$
\item for $N_-^{\infty} < N_{\ell} < N_u < N_+^{\infty} < N_0$
\item for $N_{\ell} < N_-^{\infty} < N_+^{\infty} < N_u < N_0$
\item for $N_{\ell} < N_-^{\infty} < N_u < N_+^{\infty} < N_0$
\item for $N_{\ell} < N_u < N_-^{\infty} < N_+^{\infty} < N_0$,
if $\Lambda (N_X) > {\mathcal{F}} (N_X)$
\item for $N_{\ell} < N_u < N_-^{\infty} < N_0 < N_+^{\infty}$,
if $\widetilde\Lambda (N_X,A_X) < \Phi (N_X,A_X)$
\item for $N_{\ell} < N_u < N_0 < N_-^{\infty} < N_+^{\infty}$
if $\Lambda (N_X) > {\mathcal{F}} (N_X)$ 
\begin{itemize}
    \item if $\widetilde\Lambda (N_X,A_X) > \Phi (N_X,A_X)$ and (\ref{comp_Phi_limit_A})
    \item if $\widetilde\Lambda (N_X,A_X) < \Phi (N_X,A_X)$ and not (\ref{comp_Phi_limit_A}) 
\end{itemize}
   \vspace*{-0.25cm}
\end{itemize} \\
\midrule
one intersection & \begin{itemize}
\vspace*{-0.25cm}
\item for $N_-^{\infty} < N_{\ell} < N_u < N_0 < N_+^{\infty}$,
if (\ref{comp_Phi_limit_A}) 
\item for $N_{\ell} < N_-^{\infty} < N_u < N_0 < N_+^{\infty}$,
if (\ref{comp_Phi_limit_A}) 
\item for $N_{\ell} < N_u < N_-^{\infty} < N_0 < N_+^{\infty}$, 
if $\widetilde\Lambda (N_X,A_X) > \Phi (N_X,A_X)$ and $\Lambda (N_0) > {\mathcal{F}} (N_0)$ or $\Lambda (N_0) < {\mathcal{F}} (N_0)$
 \vspace*{-0.25cm}
\end{itemize} \\
\midrule
two intersections & \begin{itemize}
\vspace*{-0.25cm}
\item for $N_{\ell} < N_u < N_-^{\infty} < N_+^{\infty} < N_0$, if $\Lambda (N_X) < {\mathcal{F}} (N_X)$
\item for $N_{\ell} < N_u < N_0 < N_-^{\infty} < N_+^{\infty}$,
if $\Lambda (N_X) < {\mathcal{F}} (N_X)$
\begin{itemize}
    \item if $\widetilde\Lambda (N_X,A_X) > \Phi (N_X,A_X)$ and not (\ref{comp_Phi_limit_A}) 
    \item if $\widetilde\Lambda (N_X,A_X) < \Phi (N_X,A_X)$ and (\ref{comp_Phi_limit_A})
\end{itemize}
 \vspace*{-0.25cm}
\end{itemize} \\
\bottomrule
\end{tabularx}
\smallskip
\caption{Intersections of $\widetilde\Lambda$ with $\Phi$ in the case (b3).}
\label{tab_b3}
\end{table}

\bibliographystyle{abbrvnat}
\bibliography{bibtex}

\end{document}